\documentclass[journal]{IEEEtran}

\usepackage{amsmath}
\usepackage{epsfig}
\usepackage{fixseminar}

\def\Ind{\mathop{\rm 1\kern -0.40em 1}\nolimits}
\def\NN{\mathop{\rm I\kern -0.20em N}\nolimits}

\newcommand{\square}{\vrule height6pt width6pt depth1pt}
\def\qed{\hfill\square}

\begin{document}
\title{Secret Key and Private Key Constructions for Simple Multiterminal Source Models}

\author{Chunxuan Ye and Prakash
Narayan

\thanks{Chunxuan Ye was with the Department of Electrical and Computer
Engineering, and the Institute for Systems Research, University of
Maryland, College Park, MD 20742. He is now with InterDigital
Communications, LLC, King of Prussia, PA 19406, USA. E-mail:
chunxuan.ye@interdigital.com.}
\thanks{Prakash Narayan is with the Department of Electrical and Computer
Engineering, and the Institute for Systems Research, University of
Maryland, College Park, MD 20742, USA. E-mail:
prakash@eng.umd.edu.}
\thanks{The work of C. Ye and P. Narayan was supported
by the National Science Foundation under Grants CCF0515124,
CCF0635271 and  CCF0830697. The material in this paper was presented in
part at the IEEE International Symposium on Information Theory,
Adelaide, Australia, Sept. 2005, and at the Information Theory and
Applications Workshop, San Diego, CA, Feb. 2006.}}

\maketitle %\spacing{2}

\begin{abstract}
We propose an approach for constructing secret and private keys
based on the long-known Slepian-Wolf code, due to Wyner, for correlated sources
connected by a virtual additive noise channel. Our work is
motivated by results of Csisz\'ar and Narayan which highlight
innate connections between secrecy generation by multiple terminals
that observe correlated source signals
and Slepian-Wolf near-lossless data compression.
Explicit procedures for such constructions and their
substantiation are provided. The performance of low density
parity check channel codes in devising a new class  of secret
keys is examined.

\vspace{0.1in}

\end{abstract}

\section{Introduction}
The problem of secrecy generation by multiple terminals, based on
their observations of separate but correlated signals followed by
public communication among themselves, has been investigated by
several authors (\cite{Mau93}, \cite{AhlCsi93}, \cite{CsiNar00},
among others). It has been shown that these terminals can generate
secrecy, namely ``common randomness'' which is kept secret from an
eavesdropper that is privy to said public communication and perhaps
also to additional ``wiretapped'' side information.

Our work is motivated by \cite{CsiNar04} which studies secrecy
generation for multiterminal ``source models'' with an arbitrary
number of terminals, each of which observes a distinct component of
a discrete memoryless multiple source (DMMS). Specifically, suppose
that $d \geq 2$ terminals observe, respectively, $n$ independent and
identically distributed (i.i.d.) repetitions of finite-valued random
variables (rvs) $X_1, \ldots, X_d$, denoted by ${\bf X}_1, \ldots,
{\bf X}_d$, where ${\bf X}_i = \left ( X_{i1}, \ldots, X_{in} \right
), \ i = 1, \ldots, d$. Thereupon, unrestricted and noiseless public
communication is allowed among the terminals. All such
communication is observed by all the terminals and by the
eavesdropper. The eavesdropper is assumed to be passive, i.e.,
unable to tamper with the public communication of the terminals. In
this framework, two models considered in \cite{CsiNar04} dealing
with a {\em secret key} (SK) and a {\em private key} (PK) are
pertinent to our work.

\noindent (i) {\em Secret key}: Suppose that all the terminals in $\{ 1, \ldots, d \}$
wish to generate a SK, i.e.,  common randomness which is concealed from the eavesdropper
with access to their public communication and which is nearly
uniformly distributed {\footnote {In \cite{CsiNar04}, a general situation is studied
in which a subset of the terminals generate a SK with the cooperation of the
remaining terminals.}}.
The largest (entropy) rate of such a SK, termed the SK capacity and
denoted by $C_{S}$, is shown in \cite{CsiNar04} to equal
\begin{equation}
C_{S}=H(X_1, \cdots ,X_d)-R_{min},
\label{e1}
\end{equation}
where
\begin{equation}
\label{e1A}
R_{min}=\min_{(R_1, \cdots , R_d)\in {\cal R}} \ \ \sum_{i=1}^dR_i,
\end{equation}
with{\footnote {Here, $\subset$ denotes a proper subset.}}
\begin{eqnarray}
\hspace{-0.1in}{\cal R}&\hspace{-0.1in}= & \hspace{-0.1in} \{(R_1,
\cdots , R_d):
\sum_{i\in B} R_i \geq \nonumber\\
& \hspace{-0.1in} & \hspace{-0.1in} H(\{X_j,\ j\in B\}|\{X_j,\  j\in
B^c\}), B\subset \{1, \cdots, d\}\}, \label{e1B}
\end{eqnarray}
where $B^c=\{1,\cdots ,d\}\backslash B$.

\vspace{0.05in}

\noindent (ii) {\em Private key}: For a given subset $A\subset \{1,
\cdots , d\}$, a PK for the terminals in $A$, private from the
terminals in $A^c$, is a SK generated by the terminals in $A$ with
the cooperation of the terminals in $A^c$, which is concealed from
an eavesdropper with access to the public interterminal
communication and also from the cooperating terminals in $A^c$ (and,
hence, private) {\footnote {A general model is considered in
\cite{CsiNar04} for privacy from a subset of $A^c$ of the cooperating
terminals.}}. The largest (entropy) rate of such a PK, termed the PK
capacity and denoted by $C_{P}(A)$, is shown in \cite{CsiNar04} to
be
\begin{eqnarray}
\hspace{-0.1in} C_{P}(A) &\hspace{-0.1in} = & \hspace{-0.1in} H(X_1, \cdots ,X_d) - H(\{ X_i, \ i \in A^c \}) - R_{min}(A) \nonumber \\
         &\hspace{-0.1in} = & \hspace{-0.1in}H(\{X_i,\ i\in A\}|\{X_i,\ i\in
A^c\})-R_{min}(A), \label{e2}
\end{eqnarray}
where
\begin{equation}
\label{e2A}
R_{min}(A)=\min_{\{R_i,\ i\in A\}\in {\cal R}(A)} \ \ \sum_{i\in A}R_i,
\end{equation}
with
\begin{eqnarray}
\hspace{-0.1in} {\cal R}(A)& \hspace{-0.1in}= & \hspace{-0.1in}
\{\{R_i,\ i\in A\}: \sum_{i\in B}
R_i\geq \nonumber\\
&\hspace{-0.1in}&  \hspace{-0.1in} H(\{X_j,\ j\in B\}|\{X_j,\  j\in
B^c\}), \ B\subset A\}. \label{e2B}
\end{eqnarray}

The expressions in (\ref{e1})--(\ref{e1B}) and
(\ref{e2})--(\ref{e2B}) afford the following interpretation
\cite{CsiNar04}. The joint entropy $H \left ( X_1, \ldots, X_d
\right )$ in (\ref{e1}) corresponds to the maximum rate of shared
common randomness -- sans secrecy constraints -- that can ever be
achieved by the terminals in $\{ 1, \ldots, d \}$ when each terminal
becomes {\em omniscient}, i.e., reconstructs all the components of
the DMMS with probability $\cong 1$ as the
observation length $n$ becomes large. Further, $R_{min}$ in
(\ref{e1A}), (\ref{e1B}) corresponds to the smallest aggregate rate
of interterminal communication that enables every terminal to
achieve omniscience \cite{CsiNar04}. Thus, from (\ref{e1}), the SK
capacity $C_{S}$, i.e., largest rate at which all the terminals in
$\{ 1, \ldots, d \}$ can generate a SK, is obtained by subtracting
from the maximum rate of shared common randomness achievable by
these terminals, viz. $H(X_1, \cdots , X_d)$, the smallest overall
rate $R_{min}$ of the (data-compressed) interterminal communication
that enables all the terminals to become omniscient. A similar
interpretation holds for the PK capacity $C_{P}(A)$ in (\ref{e2}) as
well, with the difference that the terminals in $A^c$, which
cooperate in secrecy generation and yet must not be privy to the
secrecy they help generate, can be assumed -- without loss of
generality -- to simply ``reveal'' their observations
\cite{CsiNar04}. Hence, the entropy terms in (\ref{e1}), (\ref{e1B})
are now replaced in (\ref{e2}), (\ref{e2B}) with additional
conditioning on $\{X_i,\ i\in A^c\}$. It should be noted that
$R_{min}$ and $R_{min}(A)$ are obtained as solutions to
multiterminal Slepian-Wolf (SW) (near-lossless) data compression
problems {\em not involving any secrecy constraints}.

The form of characterization of the SK and PK capacities in
(\ref{e1}) and (\ref{e2}) also suggests successive steps for generating
the corresponding keys. For instance, and loosely speaking, in
order to generate a SK,  the terminals in $\{ 1, \ldots, d \}$ first
generate common randomness (without any secrecy restrictions)
using SW-compressed interterminal communication denoted
collectively by, say, ${\bf F}$. Thus, the terminals generate rvs
$L_i = L_i ({\bf X}_i, {\bf F}), \ i \in \{ 1, \ldots, d \}$, with
$\frac{1}{n} H(L_i) > 0$, which agree with
probability $\cong 1$ for $n$ suitably large; suppressing subscripts,
let $L$ denote the
resulting ``common'' rv where $\frac{1}{n} H(L) > 0$. The second step
entails an extraction from $L$ of a SK $K = g (L)$ of entropy
rate $\frac{1}{n} H (L | {\bf F})$ by means of a suitable operation $g$
performed {\em identically}
at each terminal on the acquired common randomness $L$.
In particular, when the common randomness
acquired by the terminals corresponds to omniscience, i.e.,
$L \cong \left ( {\bf X}_1, \dots, {\bf X}_d \right )$,
and is achieved using interterminal communication
${\bf F}$ of the most parsimonious rate $\cong R_{min}$ in
(\ref{e1A}), then the corresponding SK $K = g(L)$ has the best
rate $C_S$ given by (\ref{e1}). It is important to note, however, that
as mentioned in (\cite{CsiNar04}, Section VI) and already known from
\cite{Mau93}, \cite{AhlCsi93}, neither communication by every terminal
nor omniscience is essential for generating secrecy (SK or PK) at
the best rate; for instance, the rv $L$ above need not correspond to
omniscience for the SK $K = g(L)$ to have the best possible rate in
(\ref{e1}).

A similar approach as above can be used to generate a PK
of the largest rate in (\ref{e2}).

The discussion above suggests that techniques for SW data compression
could be used to devise constructive schemes for obtaining SKs and PKs
that achieve the corresponding capacities. Further, in SW data compression,
the existence of linear encoders of rates arbitrarily close to the SW bound
has been long known \cite{Csi82}. In the special situation when the
i.i.d. sequences observed at the terminals are related to each other in
probability law through virtual discrete memoryless channels (DMCs)
characterized by independent additive noises, such linear SW encoders
can be obtained in terms of cosets of linear error correction codes for
such virtual channels, a fact first illustrated in \cite{Wyn74} for the
case of $d = 2$ terminals connected by a virtual binary symmetric
channel (BSC), and later exploited in most known linear constructions
of SW encoders (cf. e.g., \cite{AarGir02}, \cite{ColLee04}, \cite{GarZha01},
\cite{GarZho03}, \cite{HuVis04}-\cite{LanLiv04}, \cite{LivXio02}, \cite{LivLan03}, \cite{MitBaj02},
\cite{PraRam03}, \cite{SchRam04}). When the i.i.d. sequences observed
by $d = 2$ terminals are connected by an arbitrary virtual DMC, the
corresponding SW data compression can be viewed in terms of
coding for a ``semisymmetric'' channel, i.e., a channel with independent
additive noise that is defined over an enlarged alphabet \cite{HeYang06};
the case of stationary ergodic observations at the terminals is also considered
therein. These developments in SW data compression can translate into
an emergence of new constructive schemes for secrecy generation.

Motivated by these considerations, we seek to devise new
constructive schemes for secrecy generation in source models in
which SW data compression plays a central role. The main technical
contribution of this work is the following: Considering four simple
models of secrecy generation, we show how a new class of SKs and PKs
can be devised for them at rates arbitrarily close to the
corresponding capacities, relying on the SW data compression code in
\cite{Wyn74}. Additionally, we examine the performance of low density parity
check (LDPC) codes in the SW data compression step of the procedure
for secrecy generation. Preliminary results of this work have been
reported in \cite{YeNar05}, \cite{YeNar06}. In independent work
\cite{Mur04} for the case of $d=2$ terminals which is akin to but
different from ours, extraction of a SK from previously acquired
common randomness by means of a linear transformation has been
demonstrated.

In related work, SK generation for a source model with two terminals
that observe continuous-amplitude signals, has been studied in
\cite{YeRez06}, \cite{WilTse07}, \cite{Nit07}, \cite{Nit08},
\cite{YeMat10}. Furthermore, in recent years, several secrecy
generation schemes have been reported, relying  on
capacity-achieving channel codes, for ``wiretap'' secrecy models
that differ from ours. For instance, it was shown in \cite{ThaDih04}
that such a channel code can attain the secrecy capacity for any
wiretap channel. See also \cite{BloBar08}, \cite{LiaPoo08}.

%While we do not specify exactly the linear capacity-achieving channel codes used in the SW step of the procedure, these can be chosen from the class of LDPC \cite{GarZho03, LanLiv04, LivXio02, LivLan03, SchRam04} and turbo codes \cite{AarGir02, GarZha01, LiTu04, MitBaj02} that have attracted wide attention.

The paper is organized as follows. Preliminaries are contained in
Section II. In Section III, we consider four simple source models
for which we provide elementary constructive schemes for SK or PK
generation which rely on suitable SW data compression codes; the
keys thereby generated are shown to satisfy the requisite secrecy
and rate-optimality conditions in Section IV. Implementations of
these constructions using LDPC codes are illustrated in Section V
which also reports simulation results. Section VI contains closing
remarks.

%%%%%%%%%%%%%%%%%%%%%%%%%%%%%%%%%%%%%%%%%%%%%%%%%%%%%%%%%%%%%%%%%%%%%%%%%%%

\section{Preliminaries}

\subsection{Secret Key and Private Key Capacities}
Consider a DMMS with $d\geq 2$ components, with corresponding
generic rvs $X_1,\cdots ,X_d$ taking values in finite alphabets
${\cal X}_1, \cdots ,{\cal X}_d$, respectively. Let ${\bf
X}_i=(X_{i,1}, \cdots , X_{i,n})$ be $n$ i.i.d. repetitions of rv
$X_i$, $i\in {\cal D}=\{1, \cdots ,d\}$. Terminals $1, \cdots ,d$,
with respective observations ${\bf X}_1, \cdots , {\bf X}_d$,
represent the $d$ users that wish to generate a SK by means of
public communication. These terminals can communicate with each
other through broadcasts over a noiseless public channel, possibly
interactively in many rounds. In general, a communication from a
terminal is allowed to be any function of its observations, and of
all previous communication.  Let ${\bf F}$ denote collectively all
the public communication.

Given $\varepsilon>0$, the rv $K_{\cal S}$ represents an {\it
$\varepsilon$-secret key} ({\it $\varepsilon$-SK}) for the terminals
in ${\cal D}$, achieved with communication ${\bf F}$, if there exist
rvs $K_i=K_i({\bf X}_i, {\bf F})$, $i\in {\cal D}$, with $K_i$ and
$K_{\cal S}$ taking values in the same finite set ${\cal K_S}$, such
that $K_{\cal S}$ satisfies

$\bullet$ the common randomness condition
\[
\Pr\{ K_i=K_{\cal S},\ i\in {\cal D}\}\geq 1-\varepsilon;
\]

$\bullet$ the secrecy condition
\[
\frac{1}{n}I(K_{\cal S}\wedge {\bf F})\leq \varepsilon;
\]
and

$\bullet$ the uniformity condition
\[
\frac{1}{n}H(K_{\cal S})\geq \frac{1}{n}\log |{\cal K}_{\cal S}|-\varepsilon.
\]

Let $A\subset {\cal D}$ be an arbitrary subset of the terminals. The
rv $K_{\cal P}(A)$ represents an {\it $\varepsilon$-private key}
({\it $\varepsilon$-PK}) for the terminals in $A$, private from the
terminals in $A^c={\cal D}\backslash A$, achieved with communication
${\bf F}$, if there  exist rvs $K_i=K_i({\bf X}_i, {\bf F})$, $i\in
A$, with $K_i$ and $K_{\cal P}(A)$ taking values in the same finite
set ${\cal K_P}(A)$, such that $K_{\cal P}(A)$ satisfies

$\bullet$ the common randomness condition
\[
\Pr\{K_i=K_{\cal P}(A),\ i\in A\}\geq 1-\varepsilon;
\]

$\bullet$ the secrecy condition
\[
\frac{1}{n}I\left(K_{\cal P}(A)\wedge \{{\bf X}_{i},\ i\in A^c\},
{\bf F}\right)\leq \varepsilon;
\]
and

$\bullet$ the uniformity condition
\[
\frac{1}{n}H(K_{\cal P}(A))\geq \frac{1}{n}\log \left|{\cal K}_{\cal P}(A)\right|-\varepsilon.
\]

{\bf Definition 1} \cite{CsiNar04}: A nonnegative number $R$ is
called an {\it achievable SK rate} if $\varepsilon_n$-SKs $K_{\cal
S}^{(n)}$ are achievable with suitable communication (with the
number of rounds possibly depending on $n$), such that
$\varepsilon_n\rightarrow 0$ and $\frac{1}{n}H\left(K_{\cal
S}^{(n)}\right)\rightarrow R$. The largest achievable SK rate is
called the {\it SK capacity}, denoted by $C_{S}$. The PK capacity
for the terminals in $A$, denoted by $C_{P}(A)$, is similarly
defined. An achievable SK rate (resp. PK rate) will be called {\it
strongly achievable} if $\varepsilon_n$ above can be taken to vanish
exponentially in $n$. The corresponding capacities are termed {\it
strong capacities}.

Single-letter characterizations have been obtained for $C_{S}$ in
the case of $d=2$ terminals in \cite{AhlCsi93}, \cite{Mau93} and for
$d\geq 2$ terminals in \cite{CsiNar04}, given by (\ref{e1}); and for
$C_{P}(A)$ in the case of $d=3$ terminals in \cite{AhlCsi93} and for
$d\geq 3$ terminals in \cite{CsiNar04}, given by (\ref{e2}). The
proofs of the achievability parts exploit the close connection
between secrecy generation and SW data compression. Loosely
speaking, common randomness sans any secrecy restrictions is first
generated through SW-compressed interterminal communication, whereby
all the $d$ terminals acquire a (common) rv with probability $\cong
1$. In the next step, secrecy is then extracted by means of a
suitable {\em identical} operation performed at each terminal on the
acquired common randomness. When the common randomness initially
acquired by the $d$ terminals is maximal, the corresponding SK has
the best rate $C_{S}$ given by (\ref{e1}).

In this work, we consider four simple models for which we illustrate
the constructions of appropriate {\it strong} SKs or PKs.

\subsection{Linear Codes for the Binary Symmetric Channel}

The SW codes of interest will rely on the following classic result
concerning the existence of ``good'' linear channel codes for a BSC.
A BSC with crossover probability $p$, $0< p< \frac{1}{2}$, will be
denoted by BSC($p$). Let $h(p)=-p\log_2p-(1-p)\log_2(1-p)$ denote
the binary entropy function.

{\bf Lemma 1} \cite{Eli55}: For every $\varepsilon>0$, $0< p <
\frac{1}{2}$, and for all $n$  sufficiently large, there exists a
binary linear $(n,n-m)$ code for a BSC($p$), with
$m<n[h(p)+\varepsilon]$, such that the average error probability of
maximum likelihood decoding is less than $2^{-n \eta}$, for some
$\eta>0$. \qed

\subsection{Types and Typical Sequences}

The following standard facts regarding
``types'' and ``typical sequences'' and their pertinent properties
(cf. e.g., \cite{CsiKor81}) are compiled here in brief for ready
reference.

Given finite sets ${\cal X}$, ${\cal Y}$, the {\it type} of a
sequence ${\bf x}=(x_1,\cdots ,x_n)\in {\cal X}^n$, ${\cal X}$ a
finite set, is the probability mass function (pmf) $P_{\bf x}$ on
${\cal X}$ given by
\[
P_{\bf x}(a)=\frac{1}{n}|\{i:x_i=a\}|, \ \ \ a\in {\cal X},
\]
and the {\it joint type} of a pair of sequences $({\bf x}, {\bf
y})\in {\cal X}^n\times {\cal Y}^n$ is the joint pmf $P_{\bf xy}$ on
${\cal X}\times {\cal Y}$ given by
\[
P_{\bf xy}(a,b)=\frac{1}{n}|\{i:x_i=a, y_i=b\}|, \ \ \ a\in {\cal X},\ b\in {\cal Y}.
\]
The numbers of different types of sequences in ${\cal X}^n$ (resp.
${\cal X}^n\times {\cal Y}^n$) do not exceed $(n+1)^{|{\cal X}|}$
(resp. $(n+1)^{|{\cal X}||{\cal Y}|}$).

Given rvs $X$, $Y$ (taking values in ${\cal X}$, ${\cal Y}$,
respectively), with joint pmf $P_{XY}$ on ${\cal X}\times {\cal Y}$,
the set of sequences in ${\cal X}^n$ which are {\it $X$-typical with
constant $\xi$}, denoted by $T_{X,\xi}^n$, is defined as
\[
T_{X,\xi}^n\stackrel{\triangle}{=}\left\{{\bf x}\in {\cal X}^n: 2^{-n[H(X)+\xi]}\leq P_{X}^n({\bf x})\leq 2^{-n[H(X)-\xi]}\right\},
%\label{tp1}
\]
where $P_{X}^n({\bf x})\stackrel{\triangle}{=} \Pr\{{\bf X} ={\bf
x}\}$, ${\bf x}\in {\cal X}^n$; and the set of pairs of sequences in
${\cal X}^n\times {\cal Y}^n$ which are {\it $XY$-typical with
constant $\xi$}, denoted by $T_{XY,\xi}^n$, is defined as
\begin{eqnarray*}
T_{XY,\xi}^n & \hspace{-0.1in} \stackrel{\triangle}{=}&
\hspace{-0.1in} \{({\bf x}, {\bf y})\in {\cal X}^n\times {\cal Y}^n:
{\bf x}\in T_{X,\xi}^n, {\bf y}\in
T_{Y, \xi}^n, \\
&\hspace{-0.1in}& \hspace{-0.1in} 2^{-n[H(X, Y)+\xi]}\leq
P_{XY}^n({\bf x}, {\bf y})\leq 2^{-n[H(X, Y)-\xi]}\},
\end{eqnarray*}
where $P_{XY}^n({\bf x}, {\bf y})\stackrel{\triangle}{=} \Pr\{{\bf
X}={\bf x}, {\bf Y}={\bf y}\}$, ${\bf x}\in {\cal X}^n$, ${\bf y}\in
{\cal Y}^n$. It readily follows that for every $({\bf x},{\bf y})\in
T_{XY,\xi}^n$,
\[
2^{-n[H(X|Y)+2\xi]}\leq P_{X|Y}^n({\bf x}|{\bf y})\leq 2^{-n[H(X|Y)-2\xi]},
%\label{tp3}
\]
where $P_{X|Y}^n({\bf x}|{\bf y})\stackrel{\triangle}{=} \Pr\{{\bf
X}={\bf x}| {\bf Y}={\bf y}\}$, ${\bf x}\in {\cal X}^n$, ${\bf y}\in
{\cal Y}^n$.

For every ${\bf y}\in{\cal Y}^n$, the set of sequences in ${\cal
X}^n$ which are {\it $X|Y$-typical with respect to ${\bf y}$ with
constant $\xi$}, denoted by $T_{X|Y,\xi}^n({\bf y})$, is defined as
\[
T_{X|Y, \xi}^n({\bf y})\stackrel{\triangle}{=}\left\{{\bf x}\in
{\cal X}^n: ({\bf x}, {\bf y})\in T_{XY, \xi}^n\right\},
\]
with $T_{X|Y, \xi}^n({\bf y})=\phi$ if ${\bf y}\not\in T_{Y,\xi}^n$.
The following is an independent and explicit statement of the well-known fact
that the probability of a nontypical set decays to 0 exponentially
rapidly in $n$ (cf. e.g., \cite[Theorem 6.3]{Yeu08}).

{\bf Proposition 1}: Given a joint pmf $P_{XY}$ on ${\cal X}\times
{\cal Y}$ with $P_{XY}(x, y)>0$, $x\in {\cal X}$, $y\in {\cal Y}$,
for every $\xi>0$,
\begin{equation}
\sum_{{\bf x}\in T_{X,\xi}^n}P_X^n({\bf x}) \geq 1-(n+1)^{|{\cal X}|}\cdot 2^{-n \frac{\xi^2}{2\ln 2\left[\sum_{a\in {\cal X}}\log \frac{1}{P_X(a)}\right]^2}},
\label{eapp1.2}
\end{equation}
and
\begin{eqnarray}
& & \hspace{-0.58in} \sum_{({\bf x}, {\bf y})\in
T_{XY,\xi}^n}P_{XY}^n({\bf x}, {\bf
y}) \nonumber\\
&\hspace{-0.7in} \geq & \hspace{-0.3in} 1-(n+1)^{|{\cal X}||{\cal
Y}|}\cdot 2^{-n \frac{\xi^2}{2\ln 2\left[\sum_{(a, b)\in {\cal
X}\times {\cal Y}}\log \frac{1}{P_{XY}(a, b)}\right]^2}},
\label{eapp1.3}
\end{eqnarray}
for all $n\geq 1$.

\noindent{\bf Proof}: See Appendix A. \qed

%%%%%%%%%%%%%%%%%%%%%%%%%%%%%%%%%%%%%%%%%%%%%%%%%%%%%%%%%%%%%%%%%%%%%%%%%%%%

\section{Main Results}
We now present our main results on SK generation for three specific
models, and PK generation for a fourth model. The proofs of the
accompanying Theorems 1 - 4 are provided in Section IV.

\vspace{0.15in}

\noindent{\it {\large Model 1}: Let the terminals $1$ and $2$
observe, respectively, $n$ i.i.d. repetitions of the $\{0,1\}$-valued rvs
$X_1$ and $X_2$ with joint pmf
  \begin{eqnarray}
P_{X_1X_2}(x_1,x_2)&\hspace{-0.1in} =& \hspace{-0.1in}
\frac{1}{2}(1-p)\delta_{x_1x_2}+\frac{1}{2}p\
(1-\delta_{x_1x_2}),\nonumber\\
&& \hspace{1in} 0<p<\frac{1}{2}, \label{bsc}
\end{eqnarray}
with $\delta$ being the Kronecker delta function. These terminals
wish to generate a strong SK of maximum rate.}

The (strong) SK capacity for this model \cite{AhlCsi93},
\cite{CsiNar04}, \cite{Mau93}, given by (\ref{e1}), is
\[
C_{S}=I(X_1\wedge X_2)=1-h(p).
\]
We show a simple scheme for the terminals to generate a SK with rate
close to $1-h(p)$, which relies on Wyner's well-known method for SW
data compression \cite{Wyn74}. The SW problem of interest entails
terminal $2$ reconstructing the observed sequence ${\bf x}_1$ at
terminal $1$ from the SW codeword for ${\bf x}_1$ and its own
observed sequence ${\bf x}_2$.

Observe that under the given joint pmf (\ref{bsc}), ${\bf X}_2$ can
be considered as an input to a virtual BSC($p$), with corresponding
output ${\bf X}_1$, i.e., we can write
\begin{equation}
{\bf X}_1={\bf X}_2\oplus {\bf V},
\label{e2a}
\end{equation}
where ${\bf V}=(V_1,\cdots ,V_n)$ is an i.i.d. sequence of $\{0,
1\}$-valued rvs, independent of ${\bf X}_2$, and with
$\Pr\{V_i=1\}=p$, $1\leq i\leq n$.

\vspace{0.1in}

\noindent {\it (i) SW data compression} \cite{Wyn74}: Let ${\cal C}$
be a linear $(n,n-m)$ code as in Lemma 1 with parity check matrix
${\bf P}$. Both terminals know ${\cal C}$ (and ${\bf P}$). Terminal
$1$ communicates the syndrome ${\bf P}{\bf x}_1^t$ to terminal $2$. The
maximum likelihood estimate of ${\bf x}_1$ at terminal 2 is:
\[
{\hat {\bf x}_2}(1)={\bf x}_2\oplus f_{\bf P}({\bf P}{\bf x}_1^t\oplus {\bf P}{\bf x}_2^t),
\]
where $f_{\bf P}({\bf P}{\bf x}_1^t\oplus {\bf P}{\bf x}_2^t)$ is
the most likely sequence ${\bf v}\in\{0,1\}^n$ (under the pmf of
${\bf V}$ as above) with syndrome ${\bf P}{\bf v}^t={\bf P}{\bf
x}_1^t\oplus {\bf P}{\bf x}_2^t$, with $\oplus$ denoting addition
modulo 2 and $t$ denoting transposition. Note that in a standard
array corresponding to the code ${\cal C}$ above, $f_{\bf P}({\bf
P}{\bf x}_1^t\oplus {\bf P}{\bf x}_2^t)$ is simply the coset leader
of the coset with syndrome ${\bf P}{\bf x}_1^t\oplus {\bf P}{\bf
x}_2^t$. Also, ${\bf x}_1$ and ${\hat {\bf x}_2}(1)$ lie in the same
coset.

The probability of decoding error at terminal $2$ is given by
\[
\Pr\{ {\hat {\bf X}_2}(1)\neq {\bf X}_1\} = \Pr \{{\bf X}_2\oplus
f_{\bf P}({\bf P}{\bf X}_1^t\oplus {\bf P}{\bf X}_2^t)\neq {\bf
X}_1\},
\]
and it readily follows from (\ref{e2a}) that
\[
\Pr\{{\hat {\bf X}_2}(1)\neq {\bf X}_1\} = \Pr\{f_{\bf P}({\bf P}{\bf V}^t)\neq {\bf V}\}.
\]
By Lemma 1, $\Pr\{f_{\bf P}({\bf P}{\bf V}^t)\neq {\bf V}\}< 2^{-n\eta}$ for some $\eta>0$ and for all $n$ sufficiently large, so that
\[
\Pr\{{\hat {\bf X}_2}(1)= {\bf X}_1\}\geq 1-2^{-n \eta}.
\]

\noindent {\it (ii) SK construction}: Consider a (common) standard
array for ${\cal C}$ known to both terminals. Denote by ${\bf
a}_{i,j}$ the element of the $i^{th}$ row and the $j^{th}$ column in
the standard array, $1\leq i\leq 2^m$, $1\leq j\leq 2^{n-m}$.

Terminal $1$ sets $K_{1}=j_1$ if ${\bf X}_1$ equals ${\bf
a}_{i,j_1}$ in its coset $i$ in the standard array. Terminal $2$
sets $K_{2}=j_2$ if ${\hat {\bf X}_2}(1)$ equals ${\bf a}_{i,j_2}$
in the coset $i$ of the same standard array.

\vspace{0.2in}

The following theorem asserts that
$K_1$ constitutes a strong SK with rate approaching SK capacity.

{\bf Theorem 1}: Let $\varepsilon>0$ be given. Then for some
$\eta>0$ and for all $n$ sufficiently large, the pair of rvs $(K_1,
K_2)$ generated above, with (common) range ${\cal K}_1$ (say),
satisfy
\begin{equation}
\Pr\{K_{1}= K_{2}\} \geq 1-2^{-n \eta}, \label{thm1.1}
\end{equation}
\begin{equation}
I(K_1\wedge {\bf F})=0, \label{thm1.2}
\end{equation}
\begin{equation}
H(K_1)=\log |{\cal K}_1|, \label{thm1.3}
\end{equation}
and
\begin{equation}
\frac{1}{n}H(K_1)> 1-h(p)-\varepsilon. \label{thm1.4}
\end{equation}

{\it Remark}: The probability of $K_{1}$ differing from $K_{2}$
equals exactly the average error probability of maximum likelihood
decoding when ${\cal C}$ is used on a BSC($p$). Furthermore, the gap
between the rate of the generated SK and SK capacity equals the gap
between the rate of ${\cal C}$ and channel capacity.

\vspace{0.15in}

\noindent{\it {\large Model 2}: Let the terminals $1$ and $2$
observe, respectively, $n$ i.i.d. repetitions of the $\{0,1\}$-valued rvs
with joint pmf
\begin{eqnarray}
P_{X_1X_2}(0,0)&=&(1-p)(1-q),\nonumber\\
P_{X_1X_2}(0,1)&=&pq, \nonumber\\
P_{X_1X_2}(1,0)&=&p(1-q),\nonumber\\
P_{X_1X_2}(1,1)&=&q(1-p),\label{model2}
\end{eqnarray}
with $0<p< \frac{1}{2}$ and $0< q< 1$. These terminals wish to
generate a strong SK of maximum rate.}

Note that Model 1 is a special case of Model 2 for $q=\frac{1}{2}$.
We show below a scheme for the terminals to generate a SK with rate
close to the (strong) SK capacity for this model \cite{AhlCsi93},
\cite{CsiNar04}, \cite{Mau93}, which is given by (\ref{e1}) as
\[
C_{S}=I(X_1\wedge X_2)= h(p+q-2pq)-h(p).
\]

\noindent {\it (i) SW data compression}: This step is identical to
step {\it (i)} for Model 1. Note that under the given joint pmf
(\ref{model2}), ${\bf X}_1$ and ${\bf X}_2$ can be written as in
(\ref{e2a}). It follows in the same manner as for Model 1 that for
some $\eta>0$ and for all $n$ sufficiently large,
\[
\Pr\{{\hat {\bf X}_2}(1)= {\bf X}_1\}\geq 1-2^{-n\eta}.
\]

\noindent {\it (ii) SK construction}: Both terminals know the linear
$(n,n-m)$ code ${\cal C}$ as in Lemma 1, and a (common) standard
array for ${\cal C}$. Let $\{ {\bf e}_i:1\leq i\leq 2^m\}$ denote
the set of coset leaders for all the cosets of ${\cal C}$.

Denote by $A_i$ the set of sequences from $T_{X_1,\xi}^n$ in the
coset of ${\cal C}$ with coset leader ${\bf e}_i$, $1\leq i\leq
2^m$. If the number of sequences of the same type in $A_i$ is more
than $2^{n[I(X_1\wedge X_2)-\varepsilon']}$, where
$\varepsilon'>\xi+\varepsilon$ with $\varepsilon$ satisfying
$m<n[h(p)+\varepsilon]$ in Lemma 1, then collect arbitrarily
$2^{n[I(X_1\wedge X_2)-\varepsilon']}$ such sequences to compose a
subset, which we term a {\it regular subset} (as it consists of
sequences of the same type). Continue this procedure until the
number of sequences of every type in $A_i$ is less than
$2^{n[I(X_1\wedge X_2)-\varepsilon']}$. Let $N_i$ denote the number
of distinct regular subsets of $A_i$.

Enumerate (in any way) the sequences in each regular subset. Let
${\bf b}_{i,j,k}$, where $1\leq i\leq 2^m$, $1\leq j\leq N_i$,
$1\leq k\leq 2^{n[I(X_1\wedge X_2)-\varepsilon']}$, denote the
$k^{th}$ sequence of the $j^{th}$ regular subset in the $i^{th}$
coset (with coset leader ${\bf e}_i$).

Terminal $1$ sets $K_{1}=k_1$ if ${\bf X}_1$ equals ${\bf
b}_{i,j_1,k_1}$; else, $K_{1}$ is set to be uniformly distributed on
$\left\{1, \cdots, 2^{n[I(X_1\wedge X_2)-\varepsilon']}\right\}$,
independent of $({\bf X}_1, {\bf X}_2)$. Terminal $2$ sets
$K_{2}=k_2$ if ${\hat {\bf X}_2}(1)$ equals ${\bf b}_{i,j_2,k_2}$;
else, $K_{2}$ is set to be uniformly distributed on $\left\{1,
\cdots, 2^{n[I(X_1\wedge X_2)-\varepsilon']}\right\}$, independent
of $({\bf X}_1, {\bf X}_2, K_{1})$.

\vspace{0.2in}

The following theorem says that
$K_1$ constitutes a strong SK with rate approaching SK capacity.

{\bf Theorem 2}: Let $\varepsilon>0$ be given. Then for some
$\eta'=\eta'(\eta, \xi, \varepsilon, \varepsilon')>0$ and for all
$n$ sufficiently large, the pair of rvs $(K_1, K_2)$ generated
above, with range ${\cal K}_1$ (say), satisfy
\begin{equation}
\Pr\{K_{1}= K_{2}\} \geq 1-2^{-n \eta'}, \label{thm2.1}
\end{equation}
\begin{equation}
I(K_1\wedge {\bf F})=0, \label{thm2.2}
\end{equation}
\begin{equation}
H(K_1)=\log |{\cal K}_1|, \label{thm2.3}
\end{equation}
and
\begin{equation}
\frac{1}{n}H(K_1)= h(p+q-2pq)-h(p)-\varepsilon'. \label{thm2.4}
\end{equation}

\vspace{0.2in}

The next model is an instance of a {\it Markov chain on a tree} (cf.
\cite{Geo88}, \cite{CsiNar04}). Consider a tree ${\cal T}$ with
vertex set $V({\cal T})=\{1, \cdots ,d\}$ and edge set $E({\cal
T})$. For $(i,j) \in E({\cal T})$, let $B(i\leftarrow j)$ denote the
set of all vertices connected with $j$ by a path containing the edge
$(i,j)$. The rvs $X_1, \cdots , X_d$ form a {\it Markov chain on the
tree} ${\cal T}$ if for each $(i, j)\in E({\cal T})$, the
conditional pmf of $X_j$ given $\{X_l, l\in B(i\leftarrow j)\}$
depends only on $X_i$ (i.e., is conditionally independent of $\{X_l,
l\in B(i\leftarrow j)\}\backslash \{X_i\}$, conditioned on $X_i$).
Note that when ${\cal T}$ is a chain, this concept reduces to that
of a standard Markov chain.

\vspace{0.1in}

\noindent{\it {\large Model 3}: Let the terminals $1,\cdots ,d$
observe, respectively, $n$ i.i.d. repetitions of $\{0,1\}$-valued
rvs $X_1, \cdots ,X_d$ that form a Markov chain on the tree ${\cal
T}$, with joint pmf $P_{X_1\cdots X_d}$ specified as:
for $(i,j)\in E({\cal T})$,
\begin{eqnarray*}
P_{X_iX_j}(x_i,x_j)&\hspace{-0.1in}=& \hspace{-0.1in}
\frac{1}{2}(1-p_{(i, j)})\delta_{x_ix_j}+
\frac{1}{2}p_{(i, j)}\ (1-\delta_{x_ix_j}), \\
&& \hspace{1in} 0<p_{(i,j)}<\frac{1}{2},
%\label{bsc2}
\end{eqnarray*}
for $x_i, x_j\in \{0,1\}$. These $d$ terminals wish to generate a strong SK of maximum rate.}

Note that Model 1 is a special case of Model 3 for $d=2$. Without any loss of generality, let
\[
p_{max}= p_{(i^*, j^*)}=\max_{(i,j)\in E({\cal T})} p_{(i, j)} .
\]
Then, the (strong) SK capacity for this model  \cite{CsiNar04} is given by (\ref{e1}) as
\[
C_{S}=I(X_{i^*}\wedge X_{j^*})=1-h(p_{max}).
\]
We show how to extract a SK with rate close to $1-h(p_{max})$
by using an extension of the SW data compression scheme of Model 1
for reconstructing ${\bf x}_{i^*}$ at all the terminals.

\vspace{0.1in}

\noindent {\it (i) SW data compression}: Let ${\cal C}$ be the
linear $(n,n-m)$ code as in Lemma 1 for a BSC($p_{max}$), and with
parity check matrix ${\bf P}$. Each terminal $i$ communicates the
syndrome ${\bf P}{\bf x}_i^t$, $1\leq i\leq d$.

Let ${\hat {\bf x}}_{i}(j)$ denote the corresponding maximum
likelihood estimate of ${\bf x}_j$ at terminal $i$, $1\leq i\neq
j\leq d$. For a terminal $i\neq i^*$, denote by $(i_0, i_1, \cdots ,
i_r)$ the (only) path in the tree ${\cal T}$ from $i$ to $i^*$,
where $i_0=i$ and $i_r=i^*$; this terminal $i$, with the knowledge
of (${\bf x}_{i}$, ${\bf P}{\bf x}_{i_1}^t, \cdots ,{\bf P}{\bf
x}_{i_{r-1}}^t, {\bf P}{\bf x}_{i^*}^t$), forms its estimate ${\hat
{\bf x}}_{i}(i^*)$ of ${\bf x}_{i^*}$ through the following
successive maximum likelihood estimates of ${\bf x}_{i_1}, \cdots
,{\bf x}_{i_{r-1}}$:
\begin{eqnarray*}
{\hat {\bf x}}_{i}(i_1)&=&{\bf x}_{i}\oplus f_{\bf P}({\bf P}{\bf x}_{i}^t\oplus {\bf P}{\bf x}_{i_1}^t),\\
{\hat {\bf x}}_{i}(i_2)&=&{\hat {\bf x}}_i(i_1)\oplus f_{\bf P}({\bf P}{\bf x}_{i_1}^t\oplus {\bf P}{\bf x}_{i_2}^t),\\
\vdots &\vdots & \vdots\\
{\hat {\bf x}}_{i}(i_{r-1})&=&{\hat {\bf x}}_i(i_{r-2})\oplus f_{\bf P}({\bf P}{\bf x}_{i_{r-2}}^t\oplus {\bf P}{\bf x}_{i_{r-1}}^t),
\end{eqnarray*}
and finally,
\begin{equation}
{\hat {\bf x}}{i}(i^*)={\hat {\bf x}}_i(i_{r-1})\oplus f_{\bf
P}({\bf P}{\bf x}_{i_{r-1}}^t\oplus {\bf P}{\bf x}_{i^*}^t).
\label{estimate}
\end{equation}
{\bf Proposition 2}: By the successive maximum likelihood estimation
above, the estimate ${\hat {\bf X}_i}(i^*)$ at terminal $i\neq i^*$,
satisfies
\begin{equation}
\Pr\{{\hat {\bf X}_i}(i^*)= {\bf X}_{i^*}\} \geq 1- d\cdot
2^{-n\eta}, \label{e4.7a}
\end{equation}
for some $\eta>0$ and for all $n$ sufficiently large.
\label{pro4.1}

\noindent{\bf Proof}:  See Appendix B. \qed

\vspace{0.1in}

It follows directly from (\ref{e4.7a}) that for some
$\eta'=\eta'(\eta, m)>0$ and for all $n$ sufficiently large,
\[
\Pr \{{\hat {\bf X}_i}(i^*)= {\bf X}_{i^*}, 1\leq i\neq i^*\leq d
\}\geq 1-2^{-n \eta'}.
\]

\noindent {\it (ii) SK construction}: Consider a (common) standard
array for ${\cal C}$ known to all the terminals. Denote by ${\bf
a}_{l,k}$ the element of the $l^{th}$ row and the $k^{th}$ column in
the standard array, $1\leq l\leq 2^m$, $1\leq k\leq 2^{n-m}$.
Terminal $i^*$ sets $K_{i^*}=k_{i^*}$ if ${\bf X}_{i^*}$ equals
${\bf a}_{l,k_{i^*}}$ in the standard array. Terminal $i$, $1\leq
i\neq i^*\leq d$, sets $K_{i}=k_i$ if ${\hat {\bf X}_i}(i^*)$ equals
${\bf a}_{l,k_i}$ in the same standard array.

\vspace{0.2in}

The following theorem states that
$K_{i^*}$ constitutes a strong SK with rate approaching SK
capacity.

{\bf Theorem 3}: Let $\varepsilon>0$ be given. Then for some
$\eta'=\eta'(\eta, d)>0$ and for all $n$ sufficiently large, the rvs
$K_1,\cdots , K_d$ generated above, with range ${\cal K}_{i^*}$
(say), satisfy
\begin{equation}
\Pr\{K_{1}=\cdots = K_{d}\} >1- 2^{-n \eta'}, \label{thm3.1}
\end{equation}
\begin{equation}
I(K_{i^*}\wedge {\bf F})=0, \label{thm3.2}
\end{equation}
\begin{equation}
H(K_{i^*})=\log |{\cal K}_{i^*}|, \label{thm3.3}
\end{equation}
and
\begin{equation}
\frac{1}{n}H(K_{i^*})> 1-h(p_{max})-\varepsilon. \label{thm3.4}
\end{equation}

\vspace{0.15in}

\noindent{\it {\large Model 4}: Let the terminals 1, 2 and 3
observe, respectively, $n$ i.i.d. repetitions of the
$\{0,1\}$-valued rvs $X_1$, $X_2$, $X_3$, with joint pmf
$P_{X_1X_2X_3}$ given by:
\begin{eqnarray}
&&P_{X_1X_2X_3}(0,0,0)=P_{X_1X_2X_3}(0,1,1)=\frac{(1-p)(1-q)}{2},\nonumber\\
&&P_{X_1X_2X_3}(0,0,1)=P_{X_1X_2X_3}(0,1,0)=\frac{pq}{2}, \nonumber\\
&&P_{X_1X_2X_3}(1,0,0)=P_{X_1X_2X_3}(1,1,1)=\frac{p(1-q)}{2},\nonumber\\
&&P_{X_1X_2X_3}(1,0,1)=P_{X_1X_2X_3}(1,1,0)=\frac{q(1-p)}{2},\label{e2bb}
\end{eqnarray}
with $0<p< \frac{1}{2}$ and $0< q< 1$. Terminals 1 and 2 wish to
generate a strong PK of maximum rate, which is concealed from the
helper terminal 3.}

Note that under the joint pmf of $X_1$, $X_2$, $X_3$ above, we can write
\begin{equation}
{\bf X}_1={\bf X}_2\oplus {\bf X}_3 \oplus {\bf V},
\label{e2b}
\end{equation}
where ${\bf V}=(V_1, \cdots ,V_n)$ is an i.i.d. sequence of $\{0,
1\}$-valued rvs, independent of $({\bf X}_2, {\bf X}_3)$, with
$\Pr\{V_i=1\}=p$, $1\leq i\leq n$. Further, $(X_2, X_3)$ plays the
role of $(X_1, X_2)$ in Model 1 with $q$ in lieu of $p$ in the
latter.

We show below a scheme for terminals 1 and 2 to generate a PK with
rate close to (strong) PK capacity for this model \cite{AhlCsi93},
\cite{CsiNar00}, \cite{CsiNar04}, given by (\ref{e2}) as
\[
C_{P}(\{1,2\})=I(X_1\wedge X_2|X_3)= h(p+q-2pq)-h(p).
\]
The first step of this scheme entails terminal 3 simply revealing
its observations ${\bf x}_3$ to both terminals 1 and 2. Then,
Wyner's SW data compression scheme is used for reconstructing ${\bf
x}_1$ at terminal 2 from the SW codeword for ${\bf x}_1$ and its own
knowledge of ${\bf x}_2\oplus {\bf x}_3$.

\noindent {\it (i) SW data compression}: This step is identical to
step {\it (i)} for Model 1, as seen with the help of (\ref{e2b}).
Obviously,
\[
\Pr\{{\hat {\bf X}_2}(1)= {\bf X}_1\}\geq 1-2^{-n\eta},
\]
for some $\eta>0$ and for all $n$ sufficiently large.

\noindent {\it (ii) PK construction}: Suppose that terminals 1 and 2
know a linear $(n,n-m)$ code ${\cal C}$ as in Lemma 1, and a
(common) standard array for ${\cal C}$. Let $\{ {\bf e}_i:1\leq
i\leq 2^m\}$ denote the set of coset leaders for all the cosets of
${\cal C}$.

For a sequence ${\bf x}_3\in \{0,1\}^n$, denote by $A_i({\bf x}_3)$
the set of sequences from $T_{X_1|X_3,\xi}^n({\bf x}_3)$ in the
coset of ${\cal C}$ with coset leader ${\bf e}_i$, $1\leq i\leq
2^m$. If the number of sequences of the same joint type with ${\bf
x}_3$ in $A_i({\bf x}_3)$ is more than $2^{n[I(X_1\wedge
X_2|X_3)-\varepsilon']}$, where $\varepsilon'>2\xi+\varepsilon$ and
$\varepsilon$ satisfies $m<n[h(p)+\varepsilon]$ (as in Lemma 1),
then collect arbitrarily $2^{n[I(X_1\wedge X_2|X_3)-\varepsilon']}$
such sequences to compose a regular subset. Continue this procedure
until the number of sequences of every joint type with ${\bf x}_3$
in $A_i({\bf x}_3)$ is less than $2^{n[I(X_1\wedge
X_2|X_3)-\varepsilon']}$. Let $N_i({\bf x}_3)$ denote the number of
distinct regular subsets of $A_i({\bf x}_3)$.

For a given sequence ${\bf x}_3$, enumerate (in any way) the
sequences in each regular subset. Let ${\bf b}_{i,j,k}({\bf x}_3)$,
where $1\leq i\leq 2^m$, $1\leq j\leq N_i({\bf x}_3)$, $1\leq k\leq
2^{n[I(X_1\wedge X_2|X_3)-\varepsilon']}$, denote the $k^{th}$
sequence of the $j^{th}$ regular subset in the $i^{th}$ coset.

Terminal $1$ sets $K_{1}=k_1$ if ${\bf X}_1$ equals ${\bf
b}_{i,j_1,k_1}({\bf x}_3)$; else, $K_{1}$ is set to be uniformly
distributed on $\left\{1, \cdots, 2^{n[I(X_1\wedge
X_2|X_3)-\varepsilon']}\right\}$, independent of $({\bf X}_1, {\bf
X}_2,{\bf X}_3)$. Terminal 2 sets $K_{2}=k_2$ if ${\hat {\bf
X}_2}(1)$ equals ${\bf b}_{i,j_2,k_2}({\bf x}_3)$; else, $K_{2}$ is
set to be uniformly distributed on $\left\{1, \cdots,
2^{n[I(X_1\wedge X_2|X_3)-\varepsilon']}\right\}$, independent of
$({\bf X}_1, {\bf X}_2,{\bf X}_3, K_{1})$.

\vspace{0.2in}

The following theorem establishes that
$K_1$ constitutes a strong PK with rate approaching PK capacity.

{\bf Theorem 4}: Let $\varepsilon>0$ be given. Then for some
$\eta'=\eta'(\eta, \xi, \varepsilon, \varepsilon')>0$ and for all
$n$  sufficiently large, the pair of rvs $(K_1, K_2)$ generated
above, with range ${\cal K}_1$ (say), satisfy
\begin{equation}
\Pr\{K_{1}\neq K_{2}\} <2^{-n \eta'}, \label{thm4.1}
\end{equation}
\begin{equation}
I(K_1\wedge {\bf X}_3, {\bf F})=0, \label{thm4.2}
\end{equation}
\begin{equation}
H(K_1)=\log |{\cal K}_1|, \label{thm4.3}
\end{equation}
and
\begin{equation}
\frac{1}{n}H(K_1)= I(X_1\wedge X_2|X_3)-\varepsilon'.
\label{thm4.4}
\end{equation}

{\it Remark}: The PK construction scheme above applies for any joint
pmf  $P_{X_1X_2X_3}$ satisfying (\ref{e2b}), and is not restricted
to the given joint pmf  in (\ref{e2bb}).

%%%%%%%%%%%%%%%%%%%%%%%%%%%%%%%%%%%%%%%%%%%%%%%%%%%%%%%%%%%%%%%%%%%%%%%%%%%%
\section{Proofs of Theorems 1--4}

\vspace{0.1in}

\noindent{\bf Proof of Theorem 1}: It follows from the SK
construction scheme for Model 1 that
\[
\Pr\{K_{1}\neq K_{2}\}= \Pr \{{\hat {\bf X}_2(1)}\neq {\bf X}_1\}<2^{-n \eta},
\]
which is (\ref{thm1.1}). Since $X_1$ is uniformly distributed on
$\{0, 1\}$, we have for $1\leq i\leq 2^m$, $1\leq j\leq 2^{n-m}$, that
\[
\Pr \{{\bf X}_1={\bf a}_{i,j}\}=2^{-n}.
\]
Hence,
\begin{eqnarray*}
\Pr\{K_{1}=j\}&=&\sum_{i=1}^{2^m}\Pr \{{\bf X}_1= {\bf
a}_{i,j}\}\\
&=& 2^{-(n-m)}, \ \ \  1\leq j\leq 2^{n-m},
\end{eqnarray*}
i.e., $K_{1}$ is uniformly distributed on ${\cal
K}_1=\left\{1,\cdots , 2^{n-m}\right\}$, and so
\[
H(K_{1})=\log 2^{n-m}= n-m=\log |{\cal K}_1|,
\]
which is (\ref{thm1.3}). Therefore, (\ref{thm1.4}) holds since
$m<n[h(p)+\varepsilon]$.

It remains to show that $K_{1}$ satisfies (\ref{thm1.2}) with ${\bf
F}={\bf P}{\bf X}_1^t$. Let $\{{\bf e}_i, 1\leq i\leq 2^m\}$ be the
set of coset leaders for the cosets of ${\cal C}$. For $1\leq
i\leq 2^m$, $1\leq j\leq 2^{n-m}$,
\begin{eqnarray*}
\Pr \{K_{1}=j|{\bf P}{\bf X}_1^t={\bf P}{\bf e}_i^t\} & =& \frac{\Pr \{K_{1}=j, {\bf P}{\bf X}_1^t={\bf P}{\bf e}_i^t\}}{\Pr \{{\bf P}{\bf X}_1^t={\bf P}{\bf e}_i^t\}}\\
&=& \frac{\Pr \{{\bf X}_1={\bf a}_{i,j}\}}{\sum_{j'=1}^{2^{n-m}}\Pr\{{\bf X}_1={\bf a}_{i,j'}\}}\\
&=& 2^{-(n-m)}\\
&=& \Pr \{K_{1}=j\},
\end{eqnarray*}
i.e., $K_{1}$ is independent of ${\bf F}$, and so $I(K_{1}\wedge
{\bf F})=0$, establishing (\ref{thm1.2}). \qed

\vspace{0.2in}

\noindent{\bf Proof of Theorem 2}: Let ${\cal F}$ denote the union
of all regular subsets in $\bigcup_{i=1}^{2^m}A_i$. Clearly
${\cal F}\subseteq T_{X_1,\xi}^n$, so that
\begin{eqnarray}
&\hspace{-0.2in} & \hspace{-0.65in}\Pr \{{\bf X}_1\in {\cal F}\}\nonumber\\
&\hspace{-0.5in}=&\Pr \{{\bf X}_1\in T_{X_1,\xi}^n, {\bf X}_1\in {\cal F}\} \nonumber\\
&\hspace{-0.5in}=& \Pr\{{\bf X}_1\in T_{X_1,\xi}^n\}- \Pr \{{\bf
X}_1\in T_{X_1,\xi}^n\backslash {\cal F}\}. \label{e3}
\end{eqnarray}
By Proposition 1, $\Pr\{{\bf X}_1\in T_{X_1,\xi}^n\}$ goes to 1 exponentially
rapidly in $n$. We show below that
$\Pr \{{\bf X}_1\in T_{X_1,\xi}^n\backslash {\cal F}\}$ decays to 0 exponentially rapidly in $n$.

Since the number of different types of sequences in $\{0,1\}^n$ does
not exceed $(n+1)^2$, we have that
\begin{eqnarray}
\left|\{{\bf x}_1: {\bf x}_1\in T_{X_1,\xi}^n\backslash {\cal F}\}\right|
&\leq &2^m\cdot (n+1)^2 \cdot 2^{n[I(X_1\wedge X_2)-\varepsilon']} \nonumber\\
&<& (n+1)^2\cdot 2^{n[H(X_1)+\varepsilon-\varepsilon']},\nonumber
\end{eqnarray}
where the previous inequality is from
$m<n[h(p)+\varepsilon]=n[H(X_1|X_2)+\varepsilon]$.

\noindent Since $P_{X_1}^n({\bf x}_1)\leq 2^{-n[H(X_1)-\xi]}$, ${\bf x}_1\in T_{X_1,\xi}^n$, we get
\[
\Pr \{{\bf X}_1\in T_{X_1,\xi}^n\backslash {\cal F}\}< (n+1)^2\cdot 2^{-n(\varepsilon'-\xi-\varepsilon)}.
\]
Choosing $\varepsilon'>\xi+\varepsilon$, $\Pr \{{\bf X}_1\in T_{X_1,\xi}^n\backslash {\cal F}\}$ goes to 0 exponentially rapidly. Therefore, it follows from (\ref{e3}) that $\Pr \{{\bf X}_1\in {\cal F}\}$ goes to 1 exponentially rapidly in $n$, with exponent depending on $(\xi,\varepsilon, \varepsilon')$.

By the SK construction scheme for Model 2,
\begin{eqnarray*}
&&\hspace{-0.45in}\Pr \{K_{1} \neq K_{2}\} \\
&= &\Pr\{K_1\neq K_2, {\bf X}_1\in {\cal F}\}+\Pr\{K_1\neq K_2, {\bf X}_1\not\in {\cal F}\}\\
&\leq & \Pr\{ {\hat {\bf X}_2}(1)\neq {\bf X}_1, {\bf X}_1\in {\cal F}\} +\Pr \{{\bf X}_1 \not\in {\cal F}\}\\
&\leq & \Pr\{ {\hat {\bf X}_2}(1)\neq {\bf X}_1\}+ \Pr\{{\bf
X}_1\not\in {\cal F}\}.
\end{eqnarray*}
Since $\Pr \{{\hat {\bf X}_2}(1)\neq {\bf X}_1\}<2^{-n\eta}$, by the
observation in the previous paragraph, we have
\[
\Pr \{K_{1}\neq K_{2}\}<2^{-n\eta'}
\]
for some $\eta'=\eta'(\eta, \xi, \varepsilon, \varepsilon')>0$ and
for all $n$ sufficiently large, which is (\ref{thm2.1}).

Next, we shall show that $K_1$ satisfies (\ref{thm2.3}). For $1\leq
k\leq 2^{n[I(X_1\wedge X_2)-\varepsilon']}$, it is clear by choice
that
\begin{equation}
 \Pr \{K_{1}=k| {\bf X}_1\not\in {\cal F}\}= 2^{-n[I(X_1\wedge X_2)-\varepsilon']},
\label{e11}
\end{equation}
and that
\begin{eqnarray}
&\hspace{-0.2in} & \hspace{-0.6in}\Pr \{K_{1}=k| {\bf X}_1\in {\cal F}\} = \frac{\Pr\{K_{1}=k, {\bf X}_1\in {\cal F} \}}{\Pr\{{\bf X}_1\in {\cal F}\}}\nonumber\\
&\hspace{-0.2in}=& \hspace{-0.1in}\frac{\sum_{i=1}^{2^m}\sum_{j=1}^{N_i}\Pr \{{\bf X}_1={\bf b}_{i,j,k}\}}{\sum_{i=1}^{2^m}\sum_{j=1}^{N_i} 2^{n[I(X_1\wedge X_2)-\varepsilon']}\Pr \{{\bf X}_1={\bf b}_{i,j,k}\}}\label{e12}\\
&\hspace{-0.2in}=& \hspace{-0.1in}2^{-n[I(X_1\wedge
X_2)-\varepsilon']}, \label{e12g}
\end{eqnarray}
where (\ref{e12}) is due to every regular subset consisting of
sequences of the same type. From (\ref{e11}) and (\ref{e12g}),
\begin{equation}
\Pr \{K_{1}=k\}= 2^{-n[I(X_1\wedge X_2)-\varepsilon']},
\label{e12c}
\end{equation}
i.e., $K_{1}$ is uniformly distributed on ${\cal K}_1=\left\{1,
\cdots ,2^{n[I(X_1\wedge X_2)-\varepsilon']}\right\}$, with
\[
\frac{1}{n} H(K_{1})=I(X_1\wedge X_2)-\varepsilon',
\]
which is (\ref{thm2.4}).

It remains to show that $K_{1}$ satisfies (\ref{thm2.2}) with ${\bf
F}={\bf P}{\bf X}_1^t$. For $1\leq i\leq 2^m$, $1\leq k\leq
2^{n[I(X_1\wedge X_2)-\varepsilon']}$, we have
\[
\Pr \{K_{1}=k|{\bf P}{\bf X}_1^t={\bf P}{\bf e}_i^t, {\bf
X}_1\not\in {\cal F}\}=2^{-n[I(X_1\wedge X_2)-\varepsilon']}
\]
by choice, and
\begin{eqnarray}
&& \hspace{-0.8in}\Pr \{K_{1}=k| {\bf P}{\bf X}_1^t={\bf P}{\bf e}_i^t, {\bf X}_1\in {\cal F}\}\nonumber \\
&\hspace{-0.4in}=& \hspace{-0.2in}\frac{\Pr \{K_{1}=k, {\bf P}{\bf X}_1^t={\bf P}{\bf e}_i^t, {\bf X}_1\in {\cal F}\}}{\Pr \{{\bf P}{\bf X}_1^t={\bf P}{\bf e}_i^t, {\bf X}_1\in {\cal F}\}}\nonumber\\
&\hspace{-0.4in}=& \hspace{-0.2in}\frac{\sum_{j=1}^{N_i}\Pr \{{\bf X}_1={\bf b}_{i,j,k}\}}{\sum_{j=1}^{N_i}2^{n[I(X_1\wedge X_2)-\varepsilon']}\Pr \{{\bf X}_1={\bf b}_{i,j,k}\}}\nonumber\\
&\hspace{-0.4in}=& \hspace{-0.2in} 2^{-n[I(X_1\wedge
X_2)-\varepsilon']}.\nonumber
\end{eqnarray}
Hence,
\begin{eqnarray}
&& \hspace{-0.8in} \Pr \{K_{1}=k|{\bf P}{\bf X}_1^t={\bf P}{\bf e}_i^t\}\nonumber\\
&\hspace{-0.4in} =& \hspace{-0.2in} \Pr \{K_{1}=k| {\bf P}{\bf X}_1^t={\bf P}{\bf e}_i^t, {\bf X}_1\in {\cal F}\} \times \nonumber\\
&\hspace{-0.4in}& \hspace{-0.2in} \Pr \{{\bf X}_1\in {\cal F}|{\bf P}{\bf X}_1^t={\bf P}{\bf e}_i^t\} \nonumber\\
&\hspace{-0.4in}& \hspace{-0.2in}+ \Pr \{K_{1}=k|{\bf P}{\bf X}_1^t={\bf P}{\bf e}_i^t, {\bf X}_1\not\in {\cal F}\} \times \nonumber\\
&\hspace{-0.4in}& \hspace{-0.2in} \Pr \{{\bf X}_1\not\in {\cal F}|{\bf P}{\bf X}_1^t={\bf P}{\bf e}_i^t\}\nonumber\\
&\hspace{-0.4in}=& \hspace{-0.2in} 2^{-n[I(X_1\wedge X_2)-\varepsilon']}\nonumber\\
&\hspace{-0.4in}=& \hspace{-0.2in} \Pr \{K_{1}=k\}, \nonumber
\end{eqnarray}
where the previous equality follows from (\ref{e12c}). Thus, $K_{1}$
is independent of ${\bf F}$, establishing (\ref{thm2.2}). \qed

\vspace{0.2in}

\noindent{\bf Proof of Theorem 3}: Applying the same arguments used
in Theorem 1, we see that the rvs $K_1,\cdots, K_m$ satisfy
(\ref{thm3.1}), (\ref{thm3.3}) and (\ref{thm3.4}). It then remains
to show that $K_{i^*}$ satisfies
(\ref{thm3.2}) with ${\bf F}=({\bf P}{\bf X}_1^t, \cdots , {\bf
P}{\bf X}_{d}^t)$.

Under the given joint pmf $P_{X_1\cdots X_d}$, for each $i\neq i^*$, we can write
\[
{\bf X}_i={\bf X}_{i^*} \oplus {\bf V}_i,
\]
where ${\bf V}_i=(V_{i,1}, \cdots ,V_{i, n})$ is an i.i.d. sequence
of $\{0,1\}$-valued rvs. Further, ${\bf V}_i$, $1\leq i\neq i^*\leq
d$, and ${\bf X}_{i^*}$ are mutually independent. Then,
\begin{eqnarray}
&& \hspace{-0.65in} I(K_{i^*}\wedge {\bf F})\nonumber\\
&=&I(K_{i^*}\wedge \{{\bf P}{\bf X}_i^t,\ 1\leq i\leq d\}) \nonumber\\
&\leq & I(K_{i^*}\wedge {\bf P}{\bf X}_{i^*}^t, \{{\bf P}{\bf V}_i^t,\ 1\leq i\neq i^*\leq d\})\nonumber\\
&\leq & I(K_{i^*}\wedge {\bf P}{\bf X}_{i^*}^t)\nonumber\\
&& +I(K_{i^*}, {\bf P}{\bf X}_{i^*}^t\wedge \{{\bf P}{\bf V}_i^t,\
1\leq i\neq i^*\leq d\}).\label{thm3.5}
\end{eqnarray}
Clearly, the first term on the right hand side of (\ref{thm3.5}) is
zero. Since for a fixed ${\bf P}$, $(K_{i^*}, {\bf P}{\bf
X}_{i^*}^t)$ is a function of ${\bf X}_{i^*}$,
\begin{eqnarray*}
&& \hspace{-0.8in} I(K_{i^*}, {\bf P}{\bf X}_{i^*}^t\wedge \{{\bf
P}{\bf V}_i^t,\
1\leq i\neq i^*\leq d\})\\
&\leq & I({\bf X}_{i^*}\wedge \{{\bf V}_i,\ 1\leq i\neq i^*\leq
d\})=0,
\end{eqnarray*}
i.e., $K_{i^*}$ is independent of ${\bf F}$, establishing (\ref{thm3.2}). \qed

\vspace{0.2in}

\noindent{\bf Proof of Theorem 4}: For every ${\bf x}_3\in
\{0,1\}^n$, let ${\cal F}({\bf x}_3)$ denote the union of all
regular subsets in $\bigcup_{i=1}^{2^m}A_i({\bf x}_3)$. Since ${\cal
F}({\bf x}_3)\subseteq T_{X_1|X_3,\xi}^n({\bf x}_3)$,
\begin{eqnarray}
\hspace{-0.2in}\Pr \{{\bf X}_1\in {\cal F}({\bf X}_3)\}
&\hspace{-0.1in} =& \hspace{-0.1in} \Pr\{{\bf X}_1\in T_{X_1|X_3,\xi}^n({\bf X}_3)\} \nonumber\\
\hspace{-0.2in} &\hspace{-0.1in}& \hspace{-0.1in}- \Pr \{{\bf
X}_1\in T_{X_1|X_3,\xi}^n({\bf X}_3)\backslash {\cal F}({\bf
X}_3)\}.\label{e4}
\end{eqnarray}
It follows from Proposition 1 that $\Pr\{{\bf X}_1\in
T_{X_1|X_3,\xi}^n({\bf X}_3)\}$ goes to 1 exponentially rapidly in
$n$. We show below that $\Pr \{{\bf X}_1\in
T_{X_1|X_3,\xi}^n({\bf X}_3)\backslash {\cal F}({\bf X}_3)\}$ goes
to 0 exponentially rapidly in $n$.

Recall that the number of different joint types of pairs in
$\{0,1\}^n\times \{0,1\}^n$ does not exceed $(n+1)^4$. Thus,
\begin{eqnarray}
& & \hspace{-0.9in} \left|\{{\bf x}_1: {\bf x}_1\in
T_{X_1|X_3,\xi}^n({\bf
x}_3)\backslash {\cal F}({\bf x}_3)\}\right|\nonumber\\
&\hspace{-0.4in} \leq & \hspace{-0.2in} 2^m\cdot (n+1)^4 \cdot 2^{n[I(X_1\wedge X_2|X_3)-\varepsilon']} \nonumber\\
&\hspace{-0.4in}<& \hspace{-0.2in} (n+1)^4\cdot
2^{n[H(X_1|X_3)+\varepsilon-\varepsilon']},\nonumber
\end{eqnarray}
where the previous inequality is from
$m<n[h(p)+\varepsilon]=n[H(X_1|X_2,X_3)+\varepsilon]$.

Since $P_{X_1|X_3}^n({\bf x}_1|{\bf x}_3)\leq
2^{-n[H(X_1|X_3)-2\xi]}$, $({\bf x}_1, {\bf x}_3)\in T_{X_1X_3,
\xi}^n$, we get
\[
\Pr \{{\bf X}_1\in T_{X_1|X_3,\xi}^n({\bf X}_3)\backslash {\cal F}({\bf X}_3)\}< (n+1)^4\cdot 2^{-n(\varepsilon'-2\xi-\varepsilon)}.
\]
Choosing $\varepsilon'>2\xi+\varepsilon$, $\Pr \{{\bf X}_1\in T_{X_1|X_3,\xi}^n({\bf X}_3)\backslash {\cal F}({\bf X}_3)\}$ goes to 0 exponentially rapidly. Therefore, it follows from (\ref{e4}) that $\Pr \{{\bf X}_1\in {\cal F}({\bf X}_3)\}$ goes to 1 exponentially rapidly in $n$, with an exponent depending on $(\xi, \varepsilon, \varepsilon')$.

By the PK construction scheme for Model 4,
\begin{eqnarray*}
\hspace{-0.15in} && \hspace{-0.25in} \Pr \{K_{1}\neq K_{2}\} \\
\hspace{-0.15in} &=& \hspace{-0.1in} \Pr\{ K_1\neq K_2, {\bf X}_1\in {\cal F}({\bf x}_3)\} +\Pr \{K_1\neq K_2, {\bf X}_1\not\in {\cal F}({\bf x}_3)\}\\
\hspace{-0.15in}&\leq & \hspace{-0.1in} \Pr\{ {\hat {\bf X}_2}(1)\neq {\bf X}_1, {\bf X}_1\in {\cal F}({\bf x}_3)\} +\Pr \{{\bf X}_1\not\in {\cal F}({\bf x}_3)\}\\
\hspace{-0.15in} &\leq& \hspace{-0.1in} \Pr \{{\hat {\bf
X}_2}(1)\neq {\bf X}_1\}+ \Pr\{{\bf X}_1\not\in {\cal F}({\bf
X}_3)\}.
\end{eqnarray*}
Since $\Pr \{{\hat {\bf X}_2}(1)\neq {\bf X}_1\}<2^{-n\eta}$ by the
observation in the previous paragraph, we have
\[
\Pr \{K_{1}\neq K_{2}\}<2^{-n\eta'},
\]
for some $\eta'=\eta'(\eta, \xi, \varepsilon, \varepsilon')>0$ and
for all $n$ sufficiently large, which is (\ref{thm4.1}).

Next, we shall show that $K_1$ satisfies (\ref{thm4.3}). For ${\bf
x}_3\in \{0,1\}^n$ and $1\leq k\leq 2^{n[I(X_1\wedge
X_2|X_3)-\varepsilon']}$, it is clear by choice that
\[
 \Pr \{K_{1}=k| {\bf X}_1\not\in {\cal F}({\bf x}_3),{\bf X}_3={\bf x}_3\}= 2^{-n[I(X_1\wedge X_2|X_3)-\varepsilon']},
%\label{e11a}
\]
and that
\begin{eqnarray}
\hspace{-0.1in}&\hspace{-0.1in}&\hspace{-0.25in}\Pr \{K_{1}=k| {\bf X}_1\in {\cal F}({\bf x}_3), {\bf X}_3={\bf x}_3\}\nonumber\\
\hspace{-0.1in}&\hspace{-0.1in}=&\hspace{-0.1in} \frac{\Pr\{K_{1}=k, {\bf X}_1\in {\cal F}({\bf x}_3)| {\bf X}_3={\bf x}_3\}}{\Pr\{{\bf X}_1\in {\cal F}({\bf x}_3)|{\bf X}_3={\bf x}_3\}}\nonumber\\
\hspace{-0.1in}&\hspace{-0.1in}=&\hspace{-0.1in} \frac{\sum_{i=1}^{2^m}\sum_{j=1}^{N_i({\bf x}_3)}\Pr \{{\bf X}_1={\bf b}_{i,j,k}({\bf x}_3)|{\bf X}_3={\bf x}_3\}}{\sum_{i=1}^{2^m}\sum_{j=1}^{N_i({\bf x}_3)} 2^{n[I(X_1\wedge X_2|X_3)-\varepsilon']}\Pr \{{\bf X}_1={\bf b}_{i,j,k}({\bf x}_3)|{\bf X}_3={\bf x}_3\}}\nonumber\\ % \label{e12a}\\
\hspace{-0.1in}&\hspace{-0.1in}=&\hspace{-0.1in} 2^{-n[I(X_1\wedge X_2|X_3)-\varepsilon']}, \nonumber %\label{e13a}
\end{eqnarray}
where the second equality is due to every regular subset consisting
of sequences of the same joint type with ${\bf x}_3$. Therefore,
\begin{eqnarray}
&& \hspace{-0.75in} \Pr \{K_{1}=k\}=\sum_{{\bf x}_3\in \{0,1\}^n} \Pr \{K_{1}=k, {\bf X}_3={\bf x}_3\}\nonumber\\
&=& \sum_{{\bf x}_3\in \{0,1\}^n} [\Pr\{{\bf X}_1\in {\cal F}({\bf x}_3), {\bf X}_3={\bf x}_3\} \times \nonumber\\
&& \Pr \{K_{1}=k| {\bf X}_1\in {\cal F}({\bf x}_3), {\bf X}_3={\bf x}_3\}\nonumber\\
&& +\Pr\{{\bf X}_1\not\in {\cal F}({\bf x}_3), {\bf X}_3={\bf x}_3\} \times \nonumber\\
&& \Pr \{K_{1}=k| {\bf X}_1\not\in {\cal F}({\bf x}_3), {\bf X}_3={\bf x}_3\}] \nonumber\\
&=& 2^{-n[I(X_1\wedge X_2|X_3)-\varepsilon']},\label{e12b}
\end{eqnarray}
i.e., $K_{1}$ is uniformly distributed on ${\cal K}_1=\left\{1,
\cdots ,2^{n[I(X_1\wedge X_2|X_3)-\varepsilon']}\right\}$, with
\[
\frac{1}{n} H(K_{1})=I(X_1\wedge X_2|X_3)-\varepsilon',
\]
which is (\ref{thm4.4}).

It remains to show that $K_{1}$ satisfies (\ref{thm4.2}) with
$({\bf X}_3, {\bf F})=({\bf X}_3, {\bf P}{\bf X}_1^t)$. For ${\bf
x}_3\in \{0,1\}^n$, $1\leq i\leq 2^m$ and $1\leq k\leq
2^{n[I(X_1\wedge X_2|X_3)-\varepsilon']}$, we have
\begin{eqnarray*}
&&\hspace{-1.1in} \Pr \{K_{1}=k|{\bf P}{\bf X}_1^t={\bf P}{\bf
e}_i^t, {\bf X}_1\not\in {\cal F}({\bf x}_3), {\bf X}_3={\bf
x}_3\}\\
&=& 2^{-n[I(X_1\wedge X_2|X_3)-\varepsilon']}
\end{eqnarray*}
by choice, and
\begin{eqnarray}
\hspace{-0.1in}&\hspace{-0.1in}& \hspace{-0.23in}\Pr \{K_{1}=k|{\bf P}{\bf X}_1^t={\bf P}{\bf e}_i^t,{\bf X}_1\in {\cal F}({\bf x}_3), {\bf X}_3={\bf x}_3\}\nonumber\\
\hspace{-0.1in}&\hspace{-0.1in}=&\hspace{-0.1in} \frac{\Pr \{K_{1}=k, {\bf P}{\bf X}_1^t={\bf P}{\bf e}_i^t, {\bf X}_1\in {\cal F}({\bf x}_3)| {\bf X}_3={\bf x}_3\} }{\Pr \{{\bf P}{\bf X}_1^t={\bf P}{\bf e}_i^t, {\bf X}_1\in {\cal F}({\bf x}_3)| {\bf X}_3={\bf x}_3\}}\nonumber\\
\hspace{-0.1in}&\hspace{-0.1in}=&\hspace{-0.1in} \frac{\sum_{j=1}^{N_i({\bf x}_3)}\Pr \{{\bf X}_1={\bf b}_{i,j,k}({\bf x}_3)|{\bf X}_3={\bf x}_3\}}{\sum_{j=1}^{N_i({\bf x}_3)}2^{n[I(X_1\wedge X_2|X_3)-\varepsilon']}\Pr \{{\bf X}_1={\bf b}_{i,j,k}({\bf x}_3)|{\bf X}_3={\bf x}_3\}}\nonumber\\
\hspace{-0.1in}&\hspace{-0.1in}=&\hspace{-0.1in} 2^{-n[I(X_1\wedge X_2|X_3)-\varepsilon']}.\nonumber %\label{e16a}
\end{eqnarray}
Hence,
\begin{eqnarray*}
&\hspace{-0.2in}&\hspace{-1.3in} \Pr \{K_{1}=k|{\bf P}{\bf
X}_1^t={\bf P}{\bf e}_i^t,
{\bf X}_3={\bf x}_3\} \\
&\hspace{-1in} =& \hspace{-0.2in} 2^{-n[I(X_1\wedge X_2|X_3)-\varepsilon']}\\
&\hspace{-1in}=& \hspace{-0.2in} \Pr\{K_{1}=k\},
\end{eqnarray*}
where the previous equality follows from (\ref{e12b}). Thus, $K_{1}$
is independent of $({\bf X}_3, {\bf F})$, establishing
(\ref{thm4.2}). \qed
%&=& \Pr ({\bf X}_1\in {\cal F}({\bf x}_3)|{\bf P}{\bf X}_1^t={\bf P}{\bf e}_i^t, {\bf X}_3={\bf x}_3)\cdot \Pr (K_{1}=k|{\bf X}_1\in {\cal F}({\bf x}_3), {\bf P}{\bf X}_1^t={\bf P}{\bf e}_i^t, {\bf X}_3={\bf x}_3)\nonumber\\
%&& +\Pr ({\bf X}_1\not\in {\cal F}({\bf x}_3)|{\bf P}{\bf X}_1^t={\bf P}{\bf e}_i^t, {\bf X}_3={\bf x}_3)\cdot \Pr (K_{1}=k|{\bf X}_1\not\in {\cal F}({\bf x}_3), {\bf P}{\bf X}_1^t={\bf P}{\bf e}_i^t, {\bf X}_3={\bf x}_3).\nonumber\\
%&=& 2^{-n[I(X_1\wedge X_2|X_3)-\varepsilon']}\nonumber\\
%&=& \Pr (K_{1}=k),\nonumber
%\end{eqnarray}

%%%%%%%%%%%%%%%%%%%%%%%%%%%%%%%%%%%%%%%%%%%%%%%%%%%%%%%%%%%%%%%%%%%%%%%%%%%%
%\vspace{0.4in}

\section{Implementation with LDPC Codes}

We outline an implementation using LDPC codes (cf. e.g.,
\cite{Mac99}, \cite{RicUrb01}, \cite{Tan81}, \cite{Urb01}) of the
scheme for the construction of a SK for Model 1 in Section III. As
will be indicated below, similar implementations can be applied to
Models 2--4 as well.

\subsection{SK construction}
Without any loss of generality, we consider a systematic $(n,n-m)$
LDPC code ${\cal C}$ with generator matrix ${\bf G}=[{\bf I}_{n-m}\
{\bf A}]$, where ${\bf I}_{n-m}$ is an $(n-m)\times (n-m)$-identity
matrix and ${\bf A}$ is an $(n-m)\times m$-matrix. Then, the parity
check matrix for ${\cal C}$ is ${\bf P}=[{\bf A}^t\ {\bf I}_m]$,
where ${\bf I}_m$ is an $m\times m$-identity matrix. The first $n-m$
bits of every codeword in ${\cal C}$, namely the {\it information
bits}, are pairwise distinct. Further, since the coset with coset
leader ${\bf e}_i$, $1\leq i\leq 2^m$, must contain the sequence
${\bf b}_i=[{\bf 0}_{n-m}\ {\bf e}_i {\bf P}^t]$, with ${\bf
0}_{n-m}$ denoting a sequence of $n-m$ zeros, the first
($n-m$)-bit-segments of the sequences in the coset $\{{\bf
b}_i\oplus {\bf c},\ {\bf c}\in {\cal C}\}$ are pairwise distinct.

Terminal $1$ transmits the syndrome ${\bf P}{\bf x}_1^t$, whereupon
terminal $2$, knowing $({\bf x}_2, {\bf P}{\bf x}_1^t)$, applies the
belief-propagation algorithm described in \cite{LivXio02} to
estimate ${\hat {\bf x}_2}(1)$. Since the first $n-m$ bits of the
sequences in each coset are pairwise distinct, these bits can serve
as the index of a sequence in its coset. Then, terminal $1$ (resp.
2) sets $K_{1}$ (resp. $K_2$) as the first $n-m$ bits of ${\bf x}_1$
(resp. ${\hat {\bf x}_2}(1)$).

The same implementation of the SW data compression scheme above
holds for Models 2 and 4, too. It can be applied repeatedly also for
the successive estimates (\ref{estimate}) in Model 3. In Model 3,
$K_{i^*}$ (resp. $K_i$, $i\neq i^*$) is set as the first $n-m$ bits
of ${\bf x}_{i^*}$ (resp. ${\hat {\bf x}_i}(i^*)$). It should be
noted that {\it the current complexity of generating regular subsets
in Models 2 and 4 poses a hurdle for explicit efficient
constructions of a SK and a PK, respectively, for these models.}

\subsection{Simulation Results}

We provide simulation results for the tradeoff between the relative
secret key rate (i.e., the difference between the SK capacity and
the rate of the generated SK) and the rate of generating unequal SKs
at different terminals (corresponding to the bit error rate in
SK-matching), when LDPC codes are used for SK construction in Model
1.

For the purpose of comparison, three different LDPC codes were used:
(i) a $(3,4)$-regular LDPC code; (ii) a $(3,6)$-regular LDPC code; and
(iii) an irregular LDPC code with degree distribution pair (cf.
\cite{LivXio02})
\begin{eqnarray*}
\lambda(x) &=& 0.234029 x +0.212425 x^2+0.146898x^5\\
&& +0.102840x^6+0.303808x^{19},\\
\rho(x) &=& 0.71875x^7+0.28125x^8,
\end{eqnarray*}
with a common codeword length of $10^3$ bits, and upto 60 iterations
of the belief-propagation algorithm were allowed. Over $10^3$ blocks
were transmitted from terminal 1.

Simulation results are shown in Figures 1 and 2, where conditional
entropy (i.e., $H(X_1|X_2)=h(p)$) is plotted against key bit error
rate (KBER). We note that in this simulation SKs are generated at
fixed rates that are equal to the rates of the LDPC codes used. Since
for Model 1, SK capacity equals $1-h(p)$, the conditional entropy
$h(p)$ serves as an indicator of the gap between SK capacity and
the rate of the generated SK.

Figure 1 shows the performance of the $(3, 6)$-regular and the
irregular LDPC codes; Figure 2 shows the performance of the
$(3,4)$-regular LDPC code. It is seen in both figures that KBER
increases with $h(p)$. Since SK capacity decreases with increasing
$h(p)$, an increase of $h(p)$ narrows the gap between SK capacity
and the rate of the generated SK, but raises the likelihood of
generating unequal SKs at the two terminals.

It is seen from Figure 1 that the irregular LDPC code outperforms
the $(3,6)$-regular LDPC code. For instance, for a fixed crossover
probability $p=0.068$, say, and $h(p)\approx 0.3584$, the KBER for
the irregular LDPC code is as low as $10^{-5}$, while the KBER for
the $(3,6)$-regular LDPC code is only about $4\times 10^{-3}$.

        \begin{figure}
           \epsfxsize=3.3in
           \centerline{\epsffile {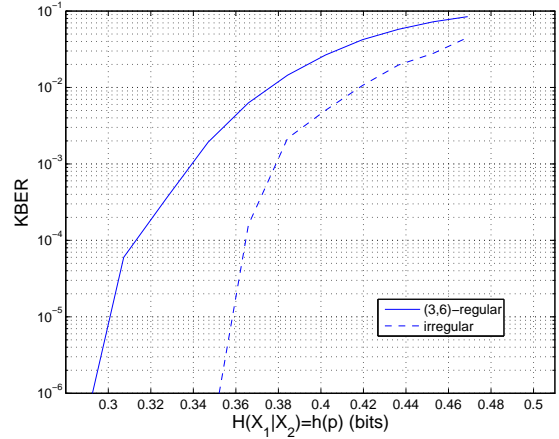}}
           \caption{Simulation results for the $(3,6)$-regular and the irregular LDPC codes.}
           \label{fig1}
         \end{figure}

        \begin{figure}
           \epsfxsize=3.3in
           \centerline{\epsffile {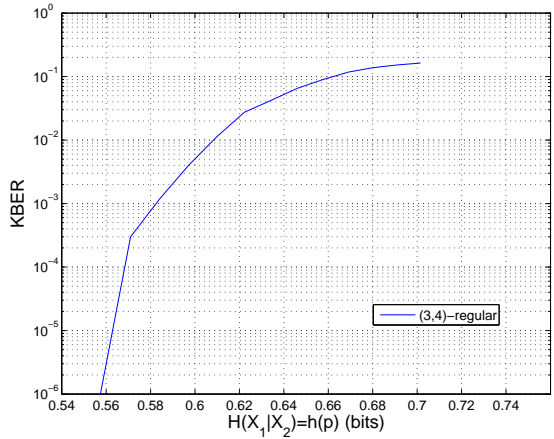}}
           \caption{Simulation results for the $(3,4)$-regular LDPC code.}
           \label{fig2}
         \end{figure}

%%%%%%%%%%%%%%%%%%%%%%%%%%%%%%%%%%%%%%%%%%%%%%%%%%%%%%%%%%%%%%%%%%%%%%%%%%%%
\section{Discussion}

We have considered four simple secrecy generation models involving
multiple terminals, and propose a new approach for constructing SKs
and PKs. This approach is based on Wyner's well-known SW data
compression code for sources connected by virtual channels with
additive independent noise.

In all the models considered in this paper, the i.i.d. sequences
observed at the different terminals possesses the following
structure: They can be described in terms of sequences at {\it pairs
of terminals} where each terminal in a pair is connected to the
other terminal by a virtual communication channel with additive
independent noise.
%
%
%are related to each other through virtual communication channels
%characterized by additive independent noises. Consider the model in
%which terminals 1 and 2, which respectively observe i.i.d.
%repetitions of the correlated $\{0,1\}$-valued rvs. $X_1$ and $X_2$
%with joint pmf $P_{X_1X_2}$, wish to generate a SK of maximal rate.
%The observations at terminal 2 can be considered as inputs to a
%virtual BSC, while the observations at terminal 1 are the
%corresponding outputs. Thus, this channel has the transition
%probability matrix $P_{X_1|X_2}$.

There are two steps in the SK construction schemes. The first step
constitutes SW data compression for the purpose of common randomness
generation at the terminals. Although the existence of linear data
compression codes with rate arbitrarily close to the SW bound has
been long known for {\it arbitrarily correlated} sources
\cite{Csi82}, constructions of such linear data compression codes
are understood in terms of the cosets of linear error-correction
codes for the virtual channel, say $P_{X_1|X_2}$, only when this
virtual channel is characterized by (independent) additive noise
\cite{Wyn74}. For instance, when two terminals are connected by a
virtual BSC $P_{X_1|X_2}$, a linear data compression code, which
attains the SW rate $H(X_1|X_2)$ for terminal 2 to reconstruct the
signal at terminal 1, is then provided by a linear channel code
which achieves the capacity of the BSC $P_{X_1|X_2}$.

When the i.i.d. sequences observed at terminals 1 and 2 are {\it
arbitrarily correlated}, the associated virtual communication
channel $P_{X_1|X_2}$ connecting them is no longer symmetric and
corresponds to a virtual channel with input-dependent noise. In this
case, while linear codes are no longer rate-optimal for the given
channel \cite{Gab67}, linear code constructions for a suitably
enlarged ``semisymmetric'' channel that are used for SW data
compression \cite{HeYang06} could pave the way for devising schemes
for SK construction.

%It is shown in \cite{Gab67} that linear codes could not achieve the
%capacity of a nonsymmetric channel in general. Therefore, it is not
%clear if the linear data compression codes that achieve the SW bound
%can be provided by the linear capacity-achieving channel codes
%anymore.

The second step in the SK construction schemes involves SK
extraction from the previously acquired CR. It has been shown
\cite{Mur04} that for the special case of a two-terminal source
model, this extraction can be accomplished by means of a linear
transformation. However, it is unknown yet whether this holds also
for a general source model with more than two terminals.

%%%%%%%%%%%%%%%%%%%%%%%%%%%%%%%%%%%%%%%%%%%%%%%%%%%%%%%%%%%%%%%%%%%%%%%%%%%%

%\begin{flushleft}
%{\Large \bf Acknowledgment}
%\end{flushleft}
%Chunxuan Ye would like to thank Dr. Alexander Barg for providing helpful references.

%%%%%%%%%%%%%%%%%%%%%%%%%%%%%%%%%%%%%%%%%%%%%%%%%%%%%%%%%%%%%%%%%%%%%%%%%%%%
\renewcommand{\theequation}{A.\arabic{equation}}
  % redefine the command that creates the equation no.
  \setcounter{equation}{0}  % reset counter
%  \section*{APPENDIX}  % use *-form to suppress numbering
\section*{Appendix A: Proof of Proposition 1}

We shall prove (\ref{eapp1.2}) here. The proof of (\ref{eapp1.3}),
which is similar, is omitted. Fix $\delta>0$ and consider the set
$T_{[P_X]_{\delta}}^n$ of sequences in ${\cal X}^n$ which are {\it
$P_{X}$-typical with constant $\delta$} (cf. \cite[p.
33]{CsiKor81}), i.e.,
\[
T_{[P_X]_{\delta}}^n=\{{\bf x}\in {\cal X}^n: \max_{a\in {\cal X}}|P_{\bf x}(a)-P_X(a)|\leq \delta\}.
\]
Since $T_{[P]_{\delta}}^n$ is the union of the sets of those types
$\tilde{P}$ of sequences in ${\cal X}^n$ that satisfy
\begin{equation}
\max_{a\in {\cal X}}|\tilde{P}(a)-P_X(a)|\leq \delta,
\label{eapp1.5}
\end{equation}
we have
\begin{eqnarray}
\hspace{-0.3in}&\hspace{-0.1in} &\hspace{-0.35in} \sum_{{\bf x}\in \left(T_{[P_X]_{\delta}}^n\right)^c} P_X^n({\bf x}) \nonumber\\
\hspace{-0.3in}&\hspace{-0.1in}= &\hspace{-0.1in} \sum_{\tilde{P}: \max_{a\in {\cal X}}|\tilde{P}(a)-P_X(a)|>\delta} P_X^n\left(\{{\bf x}: P_{\bf x}=\tilde{P}\}\right)\nonumber\\
\hspace{-0.3in}&\hspace{-0.1in}\leq &\hspace{-0.1in} (n+1)^{|{\cal
X}|} \cdot 2^{-n \min_{\tilde{P}: \min_{a\in {\cal
X}}|\tilde{P}(a)-P_X(a)|>\delta}D(\tilde{P}||P_X)}, \label{eapp1.6}
\end{eqnarray}
using the fact that $P_X^n(\{{\bf x}:P_{\bf x}=\tilde{P}\})\leq 2^{-nD(\tilde{P}||P)}$ (cf. \cite[Lemma 2.6]{CsiKor81}).

Next, by Pinsker's inequality (cf. e.g., \cite[p. 58]{CsiKor81}),
\begin{eqnarray}
D(\tilde{P}||P)&\geq & \frac{1}{2ln2}\left(\min_{a\in {\cal X}}|\tilde{P}(a)-P_X(a)|\right)^2\nonumber\\
               &\geq & \frac{\delta^2}{2ln2} ,\label{eapp1.7}
\end{eqnarray}
with the previous inequality holding for every $\tilde{P}$ in
(\ref{eapp1.5}). It follows from (\ref{eapp1.6}) and (\ref{eapp1.7})
that
\begin{equation}
\sum_{{\bf x}\in T_{[P]_{\delta}}^n}P_X^n({\bf x})\geq 1-(n+1)^{|{\cal X}|}\cdot 2^{-n \frac{\delta^2}{2ln2}}
\label{eapp1.8}
\end{equation}
for all $n\geq 1$.

Finally, observe that
\begin{equation}
T_{[P_X]_{\delta}}^n\subseteq T_{X, \xi}^n, \ \ {\rm if}\ \ \xi=\delta \left[\sum_{a\in {\cal X}}\log\frac{1}{P_X(a)}\right],
\label{eapp1.9}
\end{equation}
which is readily seen from the fact that for each ${\bf x}\in {\cal X}^n$,
\begin{eqnarray*}
&&\hspace{-0.4in} -\frac{1}{n}\log P_X^n({\bf x})-H(P_X) \\
&=& -\frac{1}{n}\log \left(2^{-n[H(P_{\bf x})+D(P_{\bf x}||P_X)]}\right)-H(P_X)\\
&=& H(P_{\bf x})+D(P_{\bf x}||P_X)-H(P_X)\\
&=& H(P_{\bf x})-H(P_{\bf x})+\sum_{a\in {\cal X}}P_{\bf x}(a)\log \frac{1}{P_X(a)}-H(P_X)\\
&=& \sum_{a\in {\cal X}}[P_{\bf x}(a)-P_X(a)]\log\frac{1}{P_X(a)}.
\end{eqnarray*}
Clearly, (\ref{eapp1.8}) and (\ref{eapp1.9}) imply (\ref{eapp1.2}). %\qed

%For an arbitrary pmf $P$ on ${\cal X}$, define a set of pmfs on ${\cal X}$ as
%\[
%E_{P,\xi}=\left\{Q:\max_{x\in {\cal X}}|Q(x)-P(x)|\leq \xi\right\}.
%\]
%Let $E_{P,\xi}^c$ denote the complement of $E_{P,\xi}$.

%It follows from Sanov's theorem (cf. \cite[p. 292]{CovTho91}) that for a generic rv $X$ with pmf $P_X$ on ${\cal X}$,
%\[
%\sum_{{\bf x}: P_{{\bf x}}\in E_{P_X,\xi}^c}P_X^n({\bf x}) \leq (n+1)^{|{\cal X}|}\cdot 2^{-n \min_{Q\in E_{P_X,\xi}^c}D(Q||P_X)}.
%\]
%By Pinsker's inequality (cf. \cite[p. 58]{CsiKor81}), for every $Q\in E_{P_X,\xi}^c$,
%\[
%D(Q||P_X)\geq \frac{\xi^2}{2\ln 2}.
%\]
%Hence,
%\[
%\sum_{{\bf x}: P_{{\bf x}}\in E_{P_X,\xi}^c}P_X^n({\bf x}) \leq (n+1)^{|{\cal X}|}\cdot 2^{-n  \frac{\xi^2}{2\ln 2}},
%\]
%or equivalently,
%\[
%\sum_{{\bf x}: P_{{\bf x}}\in E_{P_X,\xi}}P_X^n({\bf x}) \geq 1-(n+1)^{|{\cal X}|}\cdot 2^{-n  \frac{\xi^2}{2\ln 2}}.
%\]
%Since $\{{\bf x}: P_{{\bf x}}\in E_{P_X,\xi}\}\subseteq T_{X,{\bar \xi}}^n$, where ${\bar \xi}=\xi\cdot \left[\sum_{a\in {\cal X}}\log P_X(a)\right]$ (cf. \cite[p. 40]{CsiKor81}), we have
%\[
%\sum_{{\bf x}: {\bf x}\in T_{X,\xi}^n}P_X^n({\bf x}) \geq 1- (n+1)^{|{\cal X}|}\cdot 2^{-n \frac{\xi^2}{2\ln 2\left[\sum_{a\in {\cal X}}\log P_X(a)\right]^2}}.
%\]
%The second part of Proposition 1 follows from the similar arguments.

%%%%%%%%%%%%%%%%%%%%%%%%%%%%%%%%%%%%%%%%%%%%%%%%%%%%%%%%%%%%%%%%%%%%%%%%%%%
\renewcommand{\theequation}{B.\arabic{equation}}
  % redefine the command that creates the equation no.
  \setcounter{equation}{0}  % reset counter
%  \section*{APPENDIX}  % use *-form to suppress numbering
\section*{Appendix B: Proof of Proposition 2}

The proof of Proposition 2 relies on the following lemma concerning
the average error probability of maximum likelihood decoding.

A sequence ${\bf u}\in \{0,1\}^n$ is called a {\it descendent} of a sequence ${\bf v}\in \{0,1\}^n$ if $u_i=1$ implies that $v_i=1$, $1\leq i\leq n$. A subset ${\Omega}\subset \{0,1\}^n$ is called {\it quasiadmissible} if the conditions that ${\bf u}\in \Omega$ and ${\bf u}$ is a descendent of ${\bf v}$ together imply that ${\bf v}\in \Omega$.

{\bf Lemma 2} \cite{Mar74}: If $\Omega$ is a quasiadmissible subset of $\{0,1\}^n$, then for $0\leq p\leq 1$,
\[
\frac{d\mu_p(\Omega)}{dp}>0,
\]
where
\[
\mu_p({\Omega}) =\sum_{{\bf x}\in \Omega} p^{w_H({\bf
x})}(1-p)^{n-w_H({\bf x})},
\]
with $w_H({\bf x})$ denoting the Hamming weight of ${\bf x}$. \qed

\vspace{0.2in}

For a binary linear code, let ${\bf E}$ denote the set of coset
leaders. It is known (cf. \cite[Theorem 3.11]{PetWel72}) that
$\Omega'= \{0,1\}^n\backslash {\bf E}$ is a quasiadmissible subset
of $\{0,1\}^n$. If a binary linear code is used on BSC($p$), the
average error probability of maximum likelihood decoding is given by
(cf. \cite[Theorem 5.3.3]{Rom96})
\[
\mu_p(\Omega')= \sum_{{\bf x}\in \Omega'} p^{w_H({\bf
x})}(1-p)^{n-w_H({\bf x})}.
\]
Lemma 2 implies that if the same binary linear code is used on two
binary symmetric channels with different crossover probabilities,
say, $0<p_1<p_2<\frac{1}{2}$, then the average error probability of
maximum likelihood decoding for a BSC($p_1$) is strictly less than
that for a BSC($p_2$); note that a BSC($p_2$) is a degraded version
of a BSC($p_1$), being a cascade of the latter and a
BSC($\frac{p_2-p_1}{1-2p_1}$).

\vspace{0.1in}

Returning to the proof of Proposition 2, it follows from Lemma 1
that for some $\eta>0$ and for all $n$ sufficiently large,
\[
\Pr\{{\hat {\bf X}_{j^*}}(i^*)\neq {\bf X}_{i^*}\}<2^{-n\eta}.
\]
Recall that $p_{(i^*, j^*)}=\max_{(i,j)\in E({\cal T})} p_{(i, j)}$
and $(i=i_0, i_1, \cdots , i_r=i^*)$ is the path from $i$ to $i^*$.
It follows by Lemma 2 that
\[
\Pr\{{\hat {\bf X}_i}(i_1)\neq {\bf X}_{i_1}\}<\Pr\{{\hat {\bf X}_{j^*}}(i^*)\neq{\bf X}_{i^*}\}<2^{-n\eta}.
\]
Consequently,
\begin{eqnarray*}
\Pr\{{\hat {\bf X}_i}(i_2)\neq {\bf X}_{i_2}\}&\hspace{-0.1in}\leq &\hspace{-0.1in} \Pr\{{\hat {\bf X}_i}(i_2)\neq {\bf X}_{i_2}, {\hat {\bf X}_i}(i_1)\neq {\bf X}_{i_1}\}\\
                  &\hspace{-0.1in}& \hspace{-0.1in}+ \Pr\{{\hat {\bf X}_i}(i_2)\neq {\bf X}_{i_2}, {\hat {\bf X}_i}(i_1)= {\bf X}_{i_1}\}\\
                  &\hspace{-0.1in}<& \hspace{-0.1in}2\cdot 2^{-n\eta}.
\end{eqnarray*}
Continuing this procedure, we have finally that
\[
\Pr\{{\hat {\bf X}_i}(i^*)\neq{\bf X}_{i^*}\}  <r\cdot 2^{-n\eta}<d\cdot 2^{-n \eta}.
\]

%%%%%%%%%%%%%%%%%%%%%%%%%%%%%%%%%%%%%%%%%%%%%%%%%%%%%%%%%%%%%%%%%%%%%%%%%%%%

\bibliographystyle{plain}

\end{document}